\documentclass{ifacconf}

\usepackage{graphicx}      
\usepackage{natbib}        
\usepackage{amssymb}
\usepackage{amsmath}
\usepackage{algorithm}
\usepackage{algpseudocode}

\usepackage{tikz}
\usetikzlibrary{automata,positioning}
\algtext*{EndIf}
\algtext*{EndFor}
\algtext*{EndWhile}

\begin{document}
\begin{frontmatter}

\title{Lexicographic Multi-Objective Stochastic Shortest Path with Mixed Max–Sum Costs}

\thanks[footnoteinfo]{This research was supported in part by the Office of Naval Research under grant number N000142512369 and by NSF under grant number CNS-2141153.}

\author[UIUC]{Zhiquan Zhang} 
\author[Iowa]{Omar Muhammetkulyyev} 
\author[Iowa]{Tichakorn Wongpiromsarn}
\author[UIUC]{Melkior Ornik}

\address[UIUC]{University of Illinois Urbana-Champaign, 
   Urbana, IL 61820 USA (e-mail: \{zz121, mornik\}@illinois.edu).}
\address[Iowa]{Iowa State University, 
   Ames, IA 50011 USA (e-mail: \{omar99, nok\}@iastate.edu)}

\begin{abstract}                
We study the Stochastic Shortest Path (SSP) problem for autonomous systems with mixed max-sum cost aggregations under Linear Temporal Logic constraints. Classical SSP formulations rely on sum-aggregated costs, which are suitable for cumulative quantities such as time or energy but fail to capture bottleneck-style objectives such as avoiding high-risk transitions, where performance is determined by the worst single event along a trajectory. Such objectives are particularly important in safety-critical systems, where even one hazardous transition can be unacceptable. To address this limitation, we introduce max-aggregated objectives that minimize the bottleneck cost, i.e., the maximum one-step cost along a trajectory. We show that standard Bellman equations on the original state space do not apply in this setting and propose an augmented MDP with a state variable tracking the running maximum cost, together with a value iteration algorithm. We further identify a cyclic policy phenomenon, where zero-marginal-cost cycles prevent goal reaching under max-aggregation, and resolve it via a finite-horizon formulation. To handle richer task requirements, linear temporal logic specifications are translated into deterministic finite automata and combined with the system to construct a product MDP. We propose a lexicographic value iteration algorithm that handles mixed max-sum objectives under lexicographic ordering on this product MDP. Gridworld case studies demonstrate the effectiveness of the proposed framework.
\end{abstract}


\end{frontmatter}

\section{Introduction}

Stochastic transition systems are widely used in autonomous planning, such as robotic motion \citep{6385724, alterovitz2007stochastic}, communication networks \citep{7080987, 7386568}, and power grids \citep{6465655}, etc. A natural problem that arises in these settings is the stochastic shortest path (SSP) problem, which aims to find an optimal policy at each state that minimizes the expected accumulated cost. Computing the optimal policy for an SSP problem is the equivalent to solving a Markov Decision Process (MDP) with the addition of a set of goal states \citep{books/lib/BertsekasT96}.

In the traditional stochastic path problem, the cost along a path is typically accumulated using the sum operator, meaning that the total cost of a trajectory is computed by summing the costs of all individual steps along the path. This additive formulation is intuitive and mathematically convenient for modeling resources that are consumed over time, such as total traverse time consumption, energy, distance \citep{thrun2005probabilistic, bertsekas1991analysis}. However, in many safety-critical applications, simply summing costs is insufficient for obtaining the optimal policy. In these settings, the primary concern is often the magnitude of the worst-case event encountered along the trajectory rather than the total accumulated cost. A representative example is robot motion planning through constrained environments, where each state incurs a cost reflecting the local clearance or feasibility margin. A single state with extremely low clearance can make the entire trajectory infeasible, even if all other states are safe. Thus, the objective is to minimize the maximum cost encountered along the trajectory, as system failure is triggered by the worst violation rather than the cumulative effect of all states.


To address the above-mentioned scenario, it is essential to adopt a \emph{max-aggregation objective}, defined as the maximum single-step cost along the entire trajectory. Several works have studied the bottleneck cost in deterministic graphs \citep{edelkamp2005cost, slutsky2020hierarchical}. For stochastic graphs, the work of \citep{veviurko2024max} proposed a framework of max-reward reinforcement learning framework in which the agent optimizes the maximum reward encountered during an episode rather than the cumulative return. In that framework, goal-reaching is effectively incorporated into the objective, i.e., as long as the goal state has the highest reward, the optimal policy ensures reaching it, making goal reaching part of the reward design. In contrast, our setting focuses on minimizing the maximum cost along a trajectory for safety-critical objectives, while treating goal reaching as a hard constraint rather than part of the objective.

Relying only on max-aggregation can produce policies that satisfy safety thresholds but are inefficient in terms of path performance. To address this, we employ a lexicographic approach to strictly prioritize multiple objectives. While work of \citep{wray2015multi} proposed a relaxed lexicographic value iteration algorithm for optimizing over multiple objectives with lexicographic ordering, it only considered sum-aggregated costs. Furthermore, as autonomous missions become increasingly sophisticated, agents must adhere to complex high-level specifications beyond simple reachability. This motivates the integration of linear temporal logic (LTL) to formally define and enforce temporal constraints such as sequential ordering and safety rules.

In this paper, we propose a hierarchical planning algorithm for stochastic systems that considers mixed max-sum cost aggregations under LTL constraints. This formulation allows for simultaneous optimization of bottleneck minimization and cumulative cost minimization with a strict lexicographic priority structure. We identify and analyze the cyclic policy phenomenon inherent to infinite-horizon max-aggregation planning, where zero-marginal-cost cycles prevent goal reachability. To address this, we introduce a finite-horizon augmented state formulation that guarantees termination. We validate our framework through a planning scenario in a gridworld, where the state transitions are stochastic, costs are defined with lexicographic ordering, and the entire mission is defined by Linear Temporal Logic specifications.

The rest of this paper is organized as follows: Section \ref{sec:preliminary} establishes essential mathematical preliminaries and notions for our problem. Section \ref{sec:pf} formally formulates our problem. Section \ref{sec:max} provides the solution to the max-aggregation of SSP problem. Section \ref{sec:lexico} extends the max-aggregation SSP to a finite horizon version for guaranteeing goal reaching, as well as proposing a unified algorithm for solving the policy of lexicographic SSP for multiple objectives with LTL constraints. Section \ref{sec:example} validates our proposed framework with numerical examples. Section \ref{sec:conclusion} concludes this paper.

\section{Preliminaries}\label{sec:preliminary}
This section introduces essential notions related to the standard SSP problem with sum-aggregated costs, the syntax and semantics of linear temporal logic, and the construction of the product system that bridges the stochastic system dynamics with high-level specifications. 

\subsection{Stochastic Shortest Path }
\label{ssec:SSP}
A stochastic transition system \citep{teichteil2012stochastic} is a tuple $\mathcal{M} = (S, A, P, c, T)$, where
\begin{itemize}
    \item $S$ is a finite state space;
    \item $A$ is a finite action space, and $A(s) \subseteq A$ denotes the set of admissible actions at state $s\in S$;
    \item $P: S \times A\times S \rightarrow[0, 1]$ is the state transition probability function. For any $s, s' \in S$ and $a \in A(s)$, we write $P(s'|s, a)$ to denote $P(s, a, s')$, i.e., the probability that the system transitions from state $s$ to state $s'$ when action $a$ is taken;
    \item $c: S\times A\times S \rightarrow \mathbb{R}_{>0}$ is a strictly positive one-step cost function;
    \item $T \subset S$ is the set of target (terminal) states. Cost accumulation stops upon reaching any state in $T$. The first hitting time of the target set $T$ is defined as $\tau = \min\{t\ge0|s_t\in T\}$.
\end{itemize}
A stochastic transition system is essentially a Markov Decision Process (MDP) with target states. A \textit{trajectory} is an infinite sequence of states and actions $\xi = (s_0, a_0, s_1, a_1, \ldots)$, generated by repeatedly applying actions and progressing through the stochastic dynamics. A (stationary, deterministic) policy is a mapping $\pi : S \to A$ such that $\pi(s) \in A(s)$ for all $s \in S$. A policy $\pi$ is called \emph{proper} if it guarantees that the target set is reached almost surely, i.e., $\mathbb{P}_\pi(\tau < \infty | s_0) = 1$.

For an initial state $s_0$, the expected cumulative cost under a proper policy $\pi$ is defined as:
\begin{equation}
    J^\pi(s_0) = \mathbb{E}_\pi\left[\sum_{t=0}^{\tau-1}c(s_t, a_t, s_{t+1}) \Big|s_0\right].
\end{equation}
The objective of the SSP problem is to find an optimal proper policy $\pi^*$ that minimizes the expected total cost. The optimal cost function $J^*(s) = \inf_\pi J^\pi(s)$ satisfies the following Bellman optimality equation:
\begin{equation}
    J^*(s) = 
    \left\{
    \begin{array}{ll}
        \min_{a\in A(s)} \sum_{s'\in S}P(s'|s, a)(c(s, a, s') + J^*(s')), s \notin T; \\
        0, \ \ \ \ \ \ \ \ \ \ \ \ \ \ \ \ \ \ \ \ \ \ \ \ \ \ \ \ \ \ \ \ \ \ \ \ \ \ \ \ \ \ \  \ \ \ \ \ \ \ s \in T.
    \end{array}
    \right.
\end{equation}

A typical solution to this problem is through Value Iteration \citep{books/lib/BertsekasT96}. Starting from an initial function $J_0 : S \to \mathbb{R}_{\geq 0}$ (with $J_0(s)=0$ for $s\in T$), the value function is iteratively updated by the Bellman operator $\mathcal{T}$, which is defined as:
\begin{equation}
\begin{aligned}
        J_{k+1}(s) &= (\mathcal{T}J_k)(s) \\&= \min_{a\in A(s)}\sum_{s' \in S}P(s'|s, a)(c(s, a, s') + J_k(s')).
\end{aligned}
\end{equation}
To ensure the well-posedness of the SSP problem and guarantee finite-time termination, we make two standard assumptions: (i) there exists at least one proper policy, (ii) all improper policies incur infinite expected cost \citep{teichteil2012stochastic}. The latter assumption indicates that cyclic policies do not reach the goal states. Given these assumptions and the positivity of one-step costs, the Bellman operator $\mathcal{T}$ is monotonic and has a unique fixed point $J^*$, and
value iteration is guaranteed to converge to $J^*$ from any initial function that is zero on terminal states.

\subsection{Linear Temporal Logic and Deterministic Finite\\ Automata}
\label{ssec:ltl}
Beyond minimizing the accumulated costs, the system is often required to satisfy complex high-level mission specifications involving combinatorial and temporal constraints. We employ linear temporal logic (LTL) to formally specify such missions. Since we consider finite-horizon planning, we adopt finite LTL (FLTL), which is interpreted over finite sequences of states. They are often converted into Deterministic Finite Automata (DFA) to integrate FLTL specifications into the planning framework, which serve as monitors for task satisfaction \citep{Baier:2008:Principles}.

Let $AP$ be a finite set of atomic propositions. The syntax of FLTL formulas over $AP$ are defined by the following grammar
\begin{equation}
    \phi:=\top\ |\ \bot\ |\ p\ |\ \neg\phi\ |\ \phi_1 \lor \phi_2\ |\ \phi_1 \land \phi_2\  |\ X \phi\ |\ \phi_1 U \phi_2,
\end{equation}
where $p\in AP$. The Boolean constants $\top$ and $\bot$ denote ``\textit{true}'' and ``\textit{false}'' respectively. We use negation, disjunction ($\lor$), and conjunction ($\land$) to define the Boolean connectives, e.g., implication $\phi_1 \rightarrow \phi_2$ ($\neg \phi_1 \lor \phi_2$) and equivalence $\phi_1 \leftrightarrow \phi_2$ ($(\phi_1 \rightarrow \phi_2) \land (\phi_2 \rightarrow \phi_1)$). Based on the basic temporal operators ``\textit{next}'' ($X$) and ``\textit{until}'' ($U$), we introduce the derived operators ``\textit{eventually}'' and ``\textit{always}'': $F\phi = \top U \phi$, $G\phi=\neg F\neg \phi$ \citep{Baier:2008:Principles}.

The semantics of FLTL formulas are defined over finite sequences of truth assignments to the atomic propositions. A \textit{word} over $AP$ is a sequence $\sigma = \sigma_0\sigma_1\sigma_2\ldots\sigma_n$, where each $\sigma_i \subseteq AP$ represents the set of atomic propositions that hold at position $i$. We write $(\sigma, i)\models \phi$ to denote that the formula $\phi$ is satisfied at position $i$ of word $\sigma$. The satisfaction relation is defined as follows:
\begin{itemize}
    \item $(\sigma, i)\models \top$ for all $i$, and $(\sigma, i)\not\models \bot$ for all $i$;
    \item $(\sigma, i) \models p$ iff $p \in \sigma_i$ for $p \in AP$;
    \item $(\sigma, i) \models \neg \phi$ iff $(\sigma, i)\not\models \phi$;
    \item $(\sigma, i) \models \phi_1 \lor \phi_2$ iff $(\sigma, i) \models \phi_1$ or $(\sigma, i)\models \phi_2$;
    \item $(\sigma, i) \models \phi_1 \land \phi_2$ iff $(\sigma, i)\models\phi_1$ and $(\sigma, i) \models \phi_2$;
    \item $(\sigma, i) \models X \phi$ iff $i<n$ and $(\sigma, i+1)\models \phi$;
    \item $(\sigma, i) \models \phi_1 U\phi_2$ iff there exists $j$ with $i \leq j\leq n$ such that $(\sigma, j) \models \phi_2$ and for all $k$ with $i \leq k <j$, we have $(\sigma, k)\models \phi_1$.
\end{itemize}
The \emph{language} of $\phi$ is $\mathcal{L}(\phi):=\{\sigma \in (2^{AP})^* \ | \ \sigma\models\phi\}$, which collects all temporal behaviors that comply with $\phi$, where $2^{AP}$ represents the alphabet of all possible propositional valuations in a single instant, where $(2^{AP})^*$ denotes the set of all possible finite sequences of these valuations. 

Any FLTL formula $\phi$ can be translated into a deterministic finite automaton (DFA) that accepts all and only those words in $\mathcal{L}(\phi)$.
A DFA over alphabet $\Sigma$ is a tuple $\mathcal{A} = (Q, q_0, \Sigma, \delta, F)$, where $Q$ is a finite set of automaton states, $q_0 \in Q$ is the initial state, $\Sigma = 2^{AP}$ is a finite input alphabet, $\delta:Q \times \Sigma \rightarrow Q$ is the deterministic transition function, $F \subseteq Q$ is the set of accepting states. A \textit{run} of $A$ over a finite word $\sigma = \sigma_0 \sigma_1 \ldots \sigma_n$ is a state sequence $q_0q_1 \ldots q_{n+1}$ such that $q_{i+1} = \delta(q_i, \sigma_i)$ for all $i = 0, 1, \ldots, n$. The run $r$ is \textit{accepting} if the terminal state $q_{n+1}$ belongs to $F$.

\subsection{Product System Construction}
\label{ssec:product}
To incorporate an FLTL specification $\phi$ into the SSP, the stochastic transition system $\mathcal{M}$ is often combined with a DFA $\mathcal{A}$ corresponding to $\phi$ \citep{Baier:2008:Principles}. We define the product stochastic system $\mathcal{P} = (S_\mathcal{P}, A, P_{\mathcal{P}}, c_{\mathcal{P}}, T_{\mathcal{P}})$, where $S_\mathcal{P} = S \times Q$ is the product state space. The transition probability from a product state $(s, q)$ to a successor $(s', q')$ under action $a\in A$ is given by:
\begin{equation}
    P_\mathcal{P}((s, q), a, (s', q')) = 
    \left\{
    \begin{array}{ll}
        P(s, a, s'),\  q' = \delta(q, L(s')); \\
        0,\ \ \ \ \ \ \ \ \ \ \ \  {\rm otherwise}.
    \end{array}
    \right.
\end{equation}
Here, $L : S \to 2^{AP}$ is the labeling function that maps states to the atomic propositions that hold.
Thus, the stochastic evolution of the system and the deterministic progression of the automaton run synchronously. The cost function is inherited from the original SSP $c_\mathcal{P}((s, q), a, (s', q')) = c(s, a, s')$. The terminal state set $T_\mathcal{P}$ is defined as $T_\mathcal{P} = S \times F$.

Let $\tau_{\mathcal{P}}$ denote the first hitting time of the terminal set $T_{\mathcal{P}}$. By construction, a trajectory of the underlying system $\mathcal{M}$ satisfies $\phi$ if and only if the corresponding product trajectory reaches $T_{\mathcal{P}}$, i.e., $\tau_{\mathcal{P}} < \infty$. We then define the probability of satisfying $\phi$ under policy $\pi$ as $\mathbb{P}_\pi(\phi) = \mathbb{P}_\pi(\tau_{\mathcal{P}} < \infty | (s_0, q_0))$. Consequently, solving the SSP problem on $\mathcal{P}$ yields a proper policy $\pi$ that minimizes the expected accumulated cost while ensuring almost-sure satisfaction of $\phi$, i.e., $\mathbb{P}_\pi(\phi) = 1$.

\section{Problem Formulation}\label{sec:pf}
In this section, we formulate the problem of lexicographic multi-objective planning for stochastic systems subject to temporal logic constraints.

We consider a discrete stochastic system $\mathcal{M}$ as described in Section \ref{ssec:SSP} and a stochastic policy, defined as a mapping $\pi:S\times A \rightarrow [0, 1]$ satisfying $\sum_{a\in A(s)}\pi(a|s)=1$ for all $s \in S$. The set of all such policies is denoted by $\Pi$. Given an initial state $s_0 \in S$ and a policy $\pi \in \Pi$, a \emph{realization} of the stochastic transition system is a trajectory, denoted by $\xi$. In general, trajectories are infinite because the system continues evolving indefinitely. However, since we consider SSP problems where all costs and performance criteria are evaluated only up to the hitting time of reaching the goal state, it is convenient to consider the finite prefix of the trajectory that ends at the first hitting time. Thus, we define the finite trajectory as a sequence of states and actions $\xi = (s_0, a_0, s_1, a_1, \ldots, s_\tau)$, where the process terminates when $s_\tau \in T$. The probability of obtaining a specific finite trajectory $\xi$ under policy $\pi$ is obviously given by
\begin{equation}
    \mathbb{P}_\pi (\xi) = \prod_{t=0}^\tau \pi(a_t|s_t)P(s_{t+1}|s_t, a_t).
\end{equation}

Let $\Xi_\pi$ denote the set of all possible trajectories generated under policy $\pi$ starting from $s_0$. A \textit{trajectory-dependent} function is a map $\mathcal{X} : \Xi_\pi \to \mathbb{R}$, which assigns a real value $\mathcal{X}(\xi)$ to each sequence. The expectation of $\mathcal{X}$ with respect to the probability distribution induced by policy $\pi$ is then defined as 
\begin{equation}
    \mathbb{E}_\pi[\mathcal{X}] = \sum_{\xi \in \Xi_\pi} \mathcal{X}(\xi) \mathbb{P}_\pi({\xi}). 
\end{equation}

We introduce a multi-objective setting consisting of $K$ distinct objectives, indexed by the set $\mathcal{K} = \{1, \ldots, K\}$. Each objective $k\in\mathcal{K}$ is characterized by a strictly positive one-step cost function $c_k:S\times A\times S \rightarrow \mathbb{R}_{>0}$. We partition the objective set $\mathcal{K}$ into two disjoint subsets: $\mathcal{K}_{\rm sum}$ and $\mathcal{K}_{\rm max}$, such that $\mathcal{K}_{\rm sum}\cup \mathcal{K}_{\rm max} = \mathcal{K}$ and $\mathcal{K}_{\rm sum}\cap \mathcal{K}_{\rm max} = \varnothing$.
\begin{itemize}
    \item For $k \in \mathcal{K}_{\rm sum}$, the accumulated cost of a trajectory is the sum of costs incurred at each step. The expected cost is defined as:
    \begin{equation}
        J_k^\pi(s_0) := \mathbb{E}_\pi \left[\sum_{t=0}^{\tau-1}c_k(s_t, a_t, s_{t+1})\Big|s_0\right].
    \end{equation}
    \item For $k \in \mathcal{K}_{\rm max}$, the accumulated cost of a trajectory is determined by the maximum single-step cost encountered along the path. The expected cost is defined as:
    \begin{equation}
        J_k^\pi(s_0) := \mathbb{E}_\pi \left[\max_{0\leq t<\tau-1} c_k(s_t, a_t, s_{t+1})\Big|s_0\right].
    \end{equation}
\end{itemize}

Combining these $K$ objectives, we define the vector-valued cost function for a policy $\pi$ initialized at $s_0$ as $J^\pi(s_0):=(J_1^\pi(s_0), J_2^\pi(s_0), \ldots J_K^\pi(s_0))$. We employ the lexicographic ordering on $J^\pi$. Let $1, \ldots, K$ represent the priority order, where $1$ is the most critical objective and $K$ is the least critical objective. For any two policies $\pi,\pi' \in \Pi$, we define $\pi \prec_{\rm lex}\pi'$, if and only if there exists $k$ such that $J_k^\pi (s_0)< J_k^{\pi'}(s_0)$, and for $j<k$, $J_j^\pi(s_0)= J_j^{\pi'}(s_0)$. The non-strict lexicoraphical order $\preceq_{\rm lex}$ is defined in the usual way, i.e., $\pi \preceq_{\rm lex} \pi'$ if and only if $\pi \prec_{\rm lex} \pi'$ or $J_k^\pi (s_0)= J_k^{\pi'}(s_0)$ for all $k$.

Finally, the system is subject to a linear temporal logic specification $\phi$. A policy is considered admissible if the generated trajectory satisfies $\phi$.

We now formulate the problem addressed in this paper:
\begin{prob}
    Given a stochastic system $\mathcal{M}$, an FLTL formula $\phi$, a set of $K$ ordered cost functions $\{c_k\}_{k=1}^K$ and the partition $\{\mathcal{K}_{\rm sum}, \mathcal{K}_{\rm max}\}$, compute a optimal policy $\pi^* \in \Pi$ that almost surely satisfies $\phi$, i.e., $\mathbb{P}_{\pi^*}(\phi) = 1$, and lexicographically minimizes the multi-objective cost vector $J^{\pi^*}(s_0)$ among all policies that almost surely satisfy $\phi$, i.e., $\pi^* \preceq_{\rm lex} \pi, \forall \pi \in \Pi$ with $\mathbb{P}_\pi(\phi) = 1$.
\end{prob}

The above problem characterizes a hierarchical, multi-objective stochastic shortest path problem with a hybrid max-sum aggregation under FLTL specifications.

\section{Optimization of max-aggregation objectives}\label{sec:max}
We first focus on a single max-aggregated objective, corresponding to some $k \in \mathcal{K}_{\rm max}$. For simplicity, we omit the subscript $k$ in this section. We first demonstrate that the standard Bellman equation is inapplicable in this setting. Subsequently, we propose an augmented state-space formulation, the corresponding Bellman equation and a value iteration algorithm.

\subsection{Problems with standard MDP formulation}
In this subsection, we demonstrate why the standard Bellman recursion fails for max-aggregation by attempting to derive a recursive relationship for the cost function. Let $C_t$ denote the cost-to-go from time step $t$, defined as the maximum future cost: $C_t = \max\{c(s_t,a_t, s_{t+1}),c(s_{t+1},a_{t+1},\\ s_{t+2}) , \ldots\}$. By the recursive property of the maximum, we can write $C_t = \max\{c(s_t, a_t, s_{t+1}), C_{t+1}\}$.
If we define the expected cost function under policy $\pi$ directly as $J^\pi(s) = \mathbb{E}_\pi [C_t|s_t = s]$, then expanding the expectation yields:
\begin{equation}
\begin{aligned}
        J^\pi(s) &= \mathbb{E}_\pi[C_t|s_t =s]\\ &= \mathbb{E}_\pi[\max\{c(s_t, a_t, s_{t+1}), C_{t+1}\} | s_t = s]\\
        & = \sum_{a\in A}\pi(a|s)\sum_{s'\in S} \Big( P(s'|s, a)\\
        &\ \ \ \ \ \ \ \ \ \ \mathbb{E}_\pi [\max\{c(s, a, s'), C_{t+1}\} | s_{t+1} = s'] \Big).
\end{aligned}
\end{equation}

Unlike sum-aggregated costs, the maximum operator is non-linear. By Jensen's inequality, for a random variable $X$ and a constant $c$, $\mathbb{E}[\max\{X, c\}]\ge \max\{\mathbb{E}[X], c\}$. Consequently,
$\mathbb{E}_\pi [\max\{c(s, a, s'), C_{t+1}\} | s_{t+1} = s'] \geq \max\{c(s, a, s'), \mathbb{E}_\pi[C_{t+1} | s_{t+1} = s'] \}$. Thus,
\begin{equation}
    \begin{aligned}
            J^\pi(s) &\geq \sum_{a\in A}\pi(a|s)\sum_{s'\in S}P(s'|s, a)\max\{c(s, a, s'), J^\pi(s')\}.
    \end{aligned}
\end{equation}
This inequality implies that, in general, the expected max-aggregated cost at a state cannot be determined solely in terms of the expected costs of its successor states. Therefore, the principle of optimality does not hold in the original state space because optimal decisions depend on the history of the maximum cost observed so far.

\subsection{Value iteration for max-aggregation}
Similar to \citep{veviurko2024max}, we define an augmented state $s_a = (s, \lambda) \in S \times \mathbb{R}_{\ge 0}$, where $s$ denotes the current state and $\lambda$ denotes the maximum one-step cost the agent has accumulated up to the current time step. Let $J(s, \lambda)$ denote the expected max-aggregated cost starting from $(s, \lambda)$. The goal is to compute the optimal value $J^*(s_0, 0)$ from the initial state $s_0$ with no accumulated maximum.

We then derive the Bellman equation for this augmented setting. Let $C_\infty$ denote the final max-aggregated cost upon termination. Applying the law of iterated expectations $(\mathbb{E}[Z|X] = \mathbb{E}[\mathbb{E}[Z|X, Y]|X])$ \citep{sutton1998reinforcement}, we have
\begin{equation}
    \begin{aligned}
        J^\pi(s, \lambda) &= \mathbb{E}_\pi[C_\infty|s_0 = s, \lambda_0 = \lambda]\\
        &= \mathbb{E}_\pi[\mathbb{E}_\pi[C_\infty | s_0 = s, \lambda_0 = \lambda, a_0, s_1]\\&\ \ \ \ \ \ \ \  |s_0 = s, \lambda_0 = \lambda]\\
        &= \mathbb{E}_\pi[J^\pi (s_1, \max\{\lambda, c(s_0, a_0, s_1)\})\\
        &\ \ \ \ \ \ \ \ |s_0 = s, \lambda_0 = \lambda]\\
        &= \mathbb{E}_\pi[\mathbb{E}_\pi [J^\pi(s_1, \max\{\lambda, c(s, a_0, s_1)\})\\
        &\ \ \ \ \ \ \ \ |s_0 = s, \lambda_0 = \lambda, a_0]|s_0 = s, \lambda_0 = \lambda]\\
        &= \mathbb{E}_\pi [\sum_{s'} P(s'|s, a_0)J^\pi (s', \max\{\lambda, c(s, a_0, s')\})\\
        &\ \ \ \ \ \ \ \  | s_0 = s, \lambda_0 = \lambda]\\
        &= \sum_a \pi(a |(s, \lambda)) \sum_{s'}P(s'|s, a)\\
        &\ \ \ \ \ \ \ \ \ J^\pi (s', \max\{\lambda, c(s, a, s')\}).
    \end{aligned}
\end{equation}
The optimal value function $J^*$ satisfies the Bellman optimality equation
\begin{equation}
    J^*(s, \lambda) = \min_{a\in A(s)}\sum_{s'\in S} P(s'|s, a)J^*(s', \max\{\lambda, c(s, a, s')\})
\end{equation}
with boundary condition
\begin{equation}
    J^\pi(s, \lambda) = \lambda,\ \  \forall s \in T.
\end{equation}

The optimal cost can be computed via value iteration. We define the Bellman operator $\mathcal{B}$ performing on $J$ as:
\begin{equation}
    ({\mathcal{B}J})(s, \lambda) = \min_{a\in A(s)}\sum_{s' \in S}P(s'|s, a)J(s', \max\{\lambda, c(s, a, s')\}).
\end{equation}
The value iteration update rule is $J_{k+1} = \mathcal{B}J_k$. We then prove the convergence of this value iteration by monotonicity and boundedness of the Bellman operator \citep{puterman2014markov}. Assume that the one-step cost is bounded, i.e., for some $\bar c \geq 0$, $0\leq c(s, a, s')\leq\bar c, \forall s, a, s'$. We take two value functions $J_1$ and $J_2$ that satisfy $J_1(s, \lambda) \leq J_2(s, \lambda)$ for all $(s, \lambda)$. Then, 
$J_1(s', \max\{\lambda, c(s, a, s')\}) \leq J_2(s', \max\{\lambda, c(s, a, s')\})$.
Multiplying both sides by non-negative transition probabilities and summing over $s'$, we have
\begin{equation}
\begin{aligned}
        \sum_{s'}P(s'|s, a)J_1(s', \max\{\lambda, c(s, a, s')\}) 
        \leq \\\sum_{s'}P(s'|s, a)J_2(s', \max\{\lambda, c(s, a, s')\}).
\end{aligned}
\end{equation}
Thus, we can conclude that
\begin{equation}
\begin{aligned}
    (\mathcal{B}J_1)(s, \lambda) &= \min_a \sum_{s'}P(s'|s, a)J_1(s', \max\{\lambda, c(s, a, s') \})\\
    &\leq \min_a \sum_{s'}P(s'|s, a)J_2(s', \max\{\lambda, c(s, a, s')\}) \\
    &= (\mathcal{B}J_2)(s, \lambda). 
\end{aligned}
\end{equation}
In other words, the operator $\mathcal{B}$ is a monotonic operator. Besides, since every one-step cost is bounded and the operator involves a convex combination of values, the sequence $\{J_k\}$ remains bounded. Hence, the value iteration converges to the unique fixed point $J^*$.

Once $J^*$ is computed, an optimal deterministic policy is extracted as:
\begin{equation}
    \pi^*(s, \lambda) = \arg\min_{a\in A(s)}\sum_{s'}P(s'|s, a)J^*(s', \max\{\lambda, c(s, a, s')\}).
\end{equation}

This policy guarantees the minimum expected max-aggregated cost while correctly accounting for the history-dependent maximum.

\section{Finite-horizon lexicographic SSP with FLTL constraints}\label{sec:lexico}
We extend the max-aggregated SSP framework to a finite-horizon setting under FLTL specifications. We propose a unified, multi-layered value iteration algorithm that computes a policy that almost surely satisfies the FLTL specification and lexicographically optimizes an ordered set of $K$ objectives.

\subsection{Finite Horizon Formulation for Max-Aggregation}
Max-aggregation presents a fundamental challenge in infinite-horizon planning. Suppose the agent has already accumulated a cost of $\lambda$. If it traverses an edge with cost $c(s, a, s')\leq \lambda$, then the max-aggregated objective does not increase. This property creates a ``zero-marginal-cost'' region in the state space, which can induce cyclic policies where the agent loops indefinitely to avoid incurring a higher cost required to reach the target.

To prevent such behavior and ensure termination, we impose a finite planning horizon $H$. We introduce a counter $h\in\{0, 1, \ldots, H\}$ that represents the number of steps remaining before the process terminates. For the max-aggregated objective, the state is augmented to $(h, s, \lambda)$. The value function $J(h, s, \lambda)$ represents the expected cost when $h$ steps remain. The boundary conditions are modified. If $h=0$, and $s \notin T$, the agent has failed to reach the target within the predefined time limit. We assign a sufficiently large penalty cost $C_{\rm fail}\gg \bar c$ to such augmented states, where $\bar c$ is the maximal one-step cost. 

The finite-horizon Bellman equation for $h \geq 1$ is thus:
\begin{equation}
\begin{aligned}
    J(h, s, \lambda) = \min_{a\in A(s)} \sum_{s'\in S}P(s'|s, a)J(h-1, s', \\\max\{\lambda, c(s, a, s')\}).
\end{aligned}
\end{equation}

\subsection{Incorporating LTL Constraints via Product MDP}
To incorporate FLTL specification $\phi$, we translate $\phi$ into a DFA $\mathcal{A_\phi}$ as outlined in Section \ref{ssec:ltl}. For planning purposes, we extend this DFA with a rejecting sink state $q_{\rm rej}$, i.e., $\mathcal{A_\phi} = (Q \cup \{q_{\rm rej}\}, q_0, \Sigma, \delta, F)$, where $q_{\rm rej} \not\in F$. Any transition that does not follow the DFA’s transition function is redirected to $q_{\rm rej}$ and from $q_{\rm rej}$, all transitions self-loop. We then construct the usual product system $\mathcal{P}$ as defined in Section \ref{ssec:product}. For max-aggregation with finite planning horizon $H$, the product state is of the form $s_{\mathcal{P}} = (h, s, q, \lambda) \in \{0, \ldots, H\} \times S \times Q \cup \{q_{\rm rej}\} \times \mathbb{R}_{>0}$. The accepting states of the DFA define the goal region in the product. Any augmented state whose automaton component is $q_{\rm rej}$ is treated as a terminal sink with large penalty cost $C_{\rm fail} \gg \bar c$. This construction ensures that the optimal policy avoids violating the FLTL specification almost surely, because reaching $q_{\rm rej}$ incurs a maximal penalty.

\subsection{Lexicographic Value Iteration}
We now present the core algorithm: a generalized Lexicographic Value Iteration that computes a policy that minimizes the ordered set of $K$ objectives in lexicographic order. The algorithm proceeds layer by layer, where layer $k$ optimizes objective $k$ while respecting the optimality (up to a tolerance $\varepsilon$) of objectives $1,\ldots,k-1$.

First, we solve for the primary objective $J_1$ (assumed to bes the max-aggregated cost). The Bellman equation for $h \geq 1$ is:
\begin{equation}
    J_1(h,s,q,\lambda) = \min_{a\in A(s)}Q_1(h,s,q,\lambda, a),
\end{equation}
where the Q-value is given by
\begin{equation}
\begin{aligned}
        Q_1(h,s,q,\lambda,a) = \sum_{s'}P(s'|s,a)J_1(h-1, s', \delta(q,L(s')), \\\max\{\lambda, c_1(s,a,s')\}).
\end{aligned}
\end{equation}
Typically multiple actions may achieve this optimal value (or be very close to it). We identify the set of $\varepsilon$-optimal actions. We define the constrained action set $A_1^*(h,s,q,\lambda)$ as
\begin{equation}
\begin{aligned}
    A_1^*(h,s,q,\lambda) = &\Big\{a\in A(s) \ | \ Q_1(h, s, q, \lambda, a)\\
    & \ \ \ \ \ \  \leq \min_{a'} Q_1(h, s, q, \lambda, a') + \varepsilon\Big\},
    \end{aligned}
\end{equation}
where $\varepsilon\ge 0$ is a small slack parameter. 

For the subsequent objectives $k = 2, \ldots, K$, the admissible actions must remain within the action set that preserves optimality of all previous layers, i.e., $a \in A_{k-1}^*(h, s, q, \lambda)$.
Specifically, if the $k$-th objective is a sum-aggregated cost, the Bellman equation is:
\begin{equation}
    J_k(h,s,q,\lambda) = \min_{a\in A^*_{k-1}(h,s,q,\lambda)} Q_k (h,s,q,\lambda),
\end{equation}
where
\begin{equation}
\begin{aligned}
    Q_k(h,s,q,\lambda,a) = \sum_{s'}P(s'|s, a)[c_k(s,a,s') +\\ J_k(h-1,s',\delta(q,L(s')), \max\{\lambda,c(s,a,s')\})].
\end{aligned}
\end{equation}
Note that even if objective $k$ is additive, we must still track the accumulated max cost $\lambda$ in the state in a unified manner. After computing $J_k$, we refine the action set as:
\begin{equation}\small
\begin{aligned}
    A_k^*(h,s,q,\lambda) = &\Big\{a\in A_{k-1}^*(h,s,q,\lambda) \ | \\
    &Q_k(h,s,q,\lambda,a)\leq \min_{a'\in A_{k-1}^*} Q_{k}(h,s,q,\lambda,a') + \varepsilon\Big\}.
    \end{aligned}
\end{equation}

This iterative process continues until all $K$ objectives are processed. The final policy $\pi^*$ is selected from the most constrained set $A^*_K$:
\begin{equation}
    \pi^*(h,s,q,\lambda) \in \arg \min_{a\in A^*_K(h,s,q,\lambda)}Q_K(h,s,q,\lambda,a).
\end{equation}

\section{Examples}\label{sec:example}
\begin{figure}[t] \label{fig:gridworld}
\centering 
\includegraphics[width=0.47\textwidth]{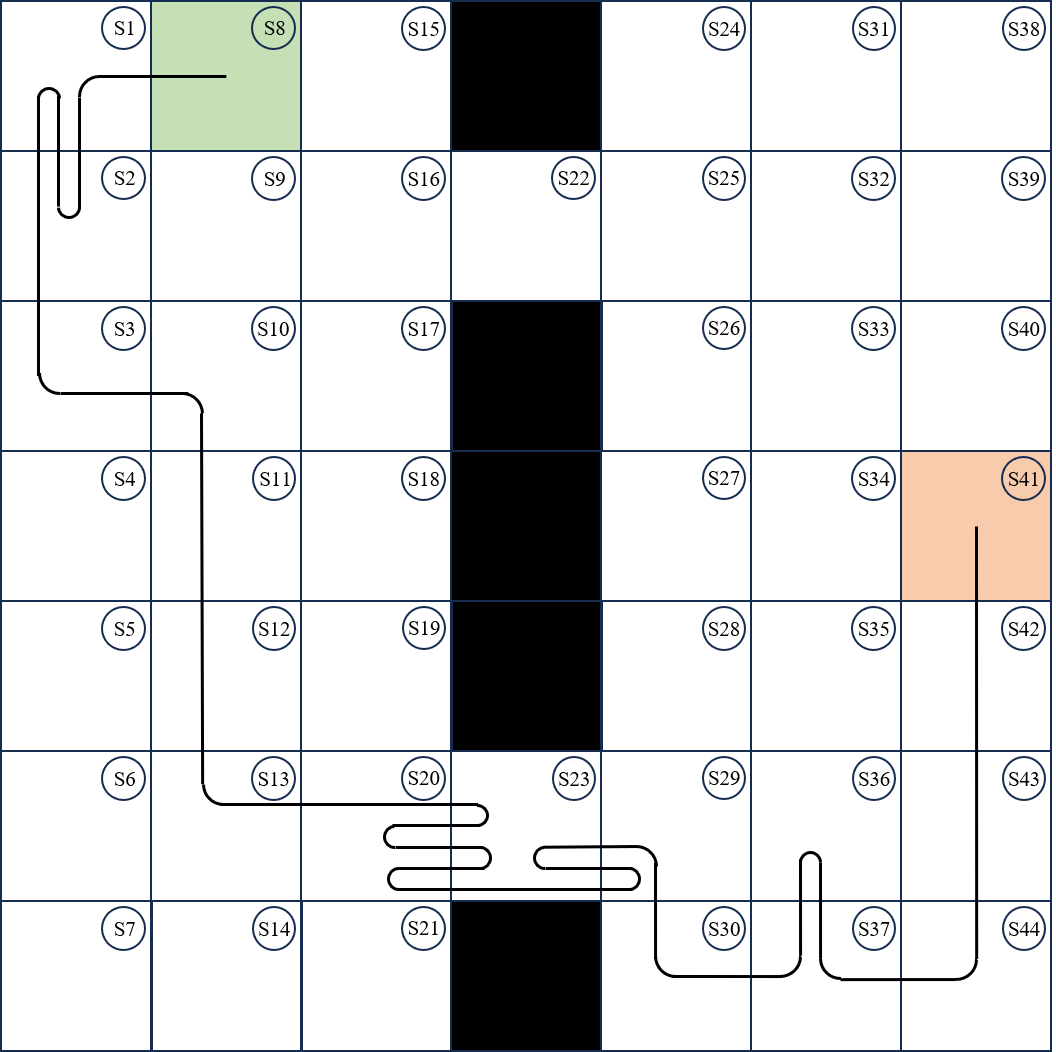} 
\caption{Visualization of state transition in the gridworld. The trajectory is generated under finite-horizon max-aggregation objective. The initial state is marked in green and the goal state is marked in red.} 
\label{Fig:fig1} 
\vspace{0.5cm} 
\includegraphics[width=0.47\textwidth]{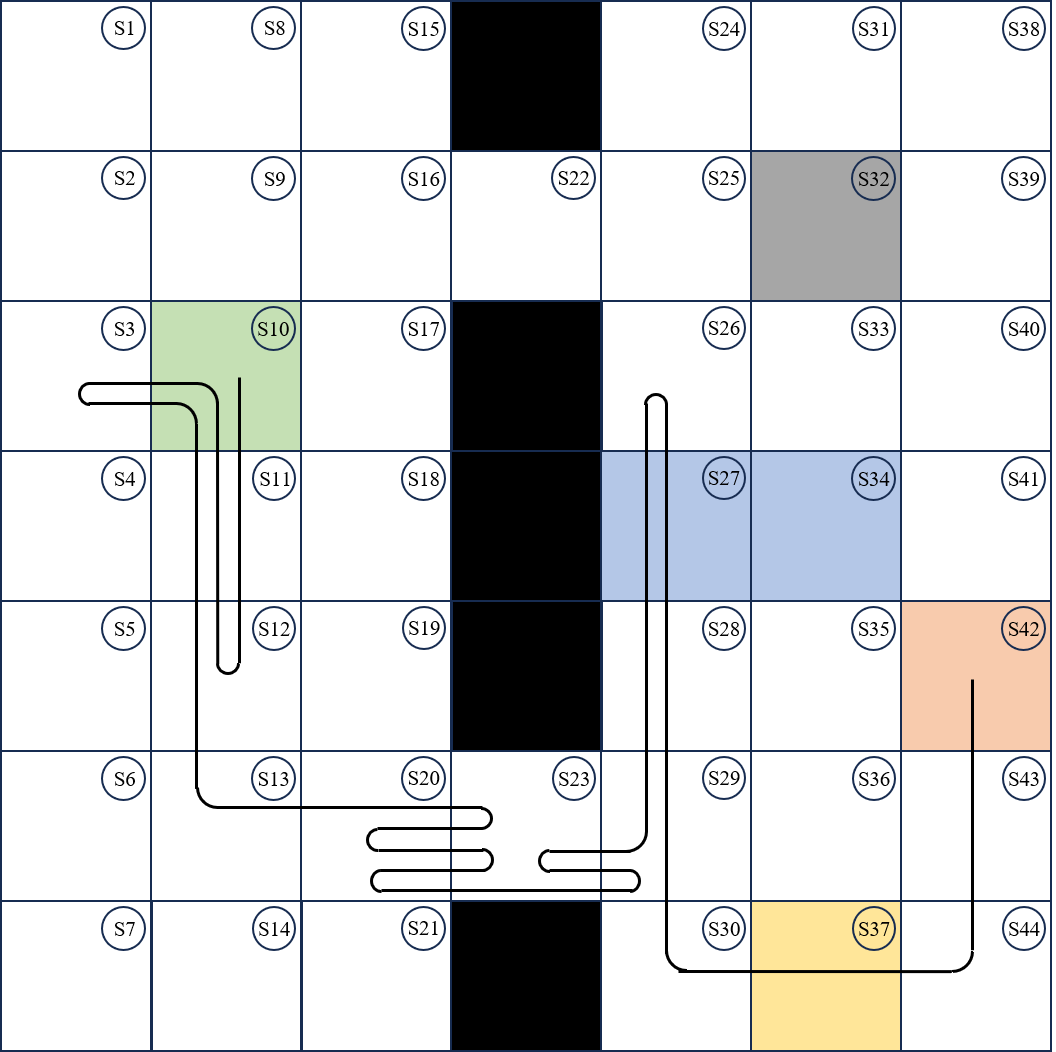} 
\caption{Visualization of state transition in the gridworld. The trajectory is generated under a prior max-aggregation objective and a secondary sum-aggregation objective with FLTL constraint. The initial state is marked in green, and the goal state is marked in red. The traversed intermediate states are marked in blue and yellow, and the obstacle is marked in grey.} 
\label{Fig:fig2} 
\end{figure}
We validate the proposed framework through two grid-world case studies, which are motivated by indoor robot navigation and similar to the settings of \citep{lacerda2019probabilistic}. The first experiment focuses on the properties of the max-aggregated objective. The second experiment evaluates the complete Lexicographic Value Iteration algorithm on a product MDP, showcasing the system's ability to satisfy Linear Temporal Logic specifications while optimizing two hierarchical costs, the first-priority cost is accumulated by max, and the second-priority cost is accumulated by sum.

The environment is a grid world constructed as shown in Fig. \ref{Fig:fig1} and Fig. \ref{Fig:fig2}. The action space consists of four directions: $A = \{U, D, L, R\}$, corresponding to up, down, left, and right, respectively. The system transitions follow a stochastic model:
\begin{itemize}
    \item Interior cells (4 neighbors): The agent moves to the intended cell with probability 0.7, and slips into any of the other three adjacent cells with probability 0.1 each.
    \item Boundary cells (3 neighbors): The success probability is 0.8, with 0.1 probability of slipping into each of the two other neighbors.
    \item Corner cells (2 neighbors): The success probability is 0.9, with 0.1 probability of slipping into the alternative neighbor.
\end{itemize}
The one-step max-aggregated cost is set to be 20 for all states, with the exception of two critical gateway states: $s22$ has a cost of $90$, and $s23$ has a cost of 30.

We first evaluate the SSP problem with a single max-aggregated objective to demonstrate the distinct behavior of the proposed cost metric. The agent starts at $s8$ and aims to reach the goal at $s41$. The finite horizon is set to $H=200$, with a failure penalty cost of $10^6$. The resulting trajectory is shown in Fig. \ref{Fig:fig1}. The action sequence is $(L, D, D, D, D, D, D, D, D, R, R, R, R, R, R, R, D, R, D, R,\\ R, R, R, U, U, U)$, and the state sequence is $(s8, s1, s2, s1,\\ s2, s3, s10, s11, s12, s13, s20, s23, s20, s23, s20, s23, s29, s23,\\ s29, s23, s29, s30, s37, s36, s37, s44, s43, s42, s41)$. Note that a critical decision point arises when the agent transitions from the left part to the right part of the grid-world. It must choose between passing through $s22$ or $s23$. A standard sum-aggregated policy tends to minimize the total accumulated cost. If the path through $s22$ had less costs, a sum-minimizing agent might choose $s22$ to pass. In contrast, our max-aggregated policy identifies $s22$ as a severe bottleneck. The agent explicitly selects the path through $s23$, accepting a potentially longer route to ensure the maximum single-step cost never exceeds 30.

In the second experiment, we apply a two-phase Lexicographic Value Iteration to solve a problem with FLTL specification. With almost-sure that the trajectory satisfies a FLTL specification, the primary objective $(J_1)$ is to minimize the max-aggregated cost, and the secondary objective $(J_2)$ is to minimize the sum-aggregated cost. 

\begin{figure}[ht]
    \centering

    \begin{tikzpicture}[
    >=stealth,
    auto,
    node distance=2.2cm,
    on grid,
    every state/.style={minimum size=14pt}
    ]
      \node[state, initial](q0){$q_0$};
      \node[state](q1)[right=of q0]{$q_1$};
      \node[state](q2)[right=of q1] {$q_2$};
      \node[state, accepting]         (q3) [right=of q2] {$q_3$};
      \node[state] (qd) [below=3.0cm of q0] {$q_{\text{dead}}$};

      \path[->]
        (q0) edge[loop above] node[align=center, above=3pt] {other} ()
        (q1) edge[loop above] node[align=center, above=3pt] {$\alpha$, other} ()
        (q2) edge[loop above] node[align=center, above=3pt] {$\alpha, \beta$, other} ()

        (qd) edge[loop below] node[align=center, below=3pt] {$\alpha, \beta, \gamma$,\\bad, other} ();

      \path[->]
        (q0) edge node[above=2pt] {$\alpha$} (q1)
        (q1) edge node[above=2pt] {$\beta$}  (q2)
        (q2) edge node[above=2pt] {$\gamma$} (q3);

      \path[->]
        (q0) edge[bend left=25]  node[above, sloped, pos=0.35, rotate=180] {$\beta, \gamma$, bad} (qd)
        (q1) edge[bend left=20]  node[above, sloped, pos=0.35] {$\gamma$, bad} (qd)
        (q2) edge[bend left=15]  node[above, sloped, pos=0.35] {bad} (qd)
        (q3) edge[bend left=10]  node[above, sloped, pos=0.35] {bad} (qd);
        s
    \end{tikzpicture}
    \caption{Visualisation of the automaton translated by LTL specification \eqref{eq:LTL}, where $\alpha$, $\beta$, $\gamma$ denote transitions performed by reaching states $\{s27, s34\}$,  $s37$, and $s41$, respectively. \textit{bad} denotes states $s32$, \textit{other} denotes the other states.}
    \label{fig:LTL_automaton}
\end{figure}
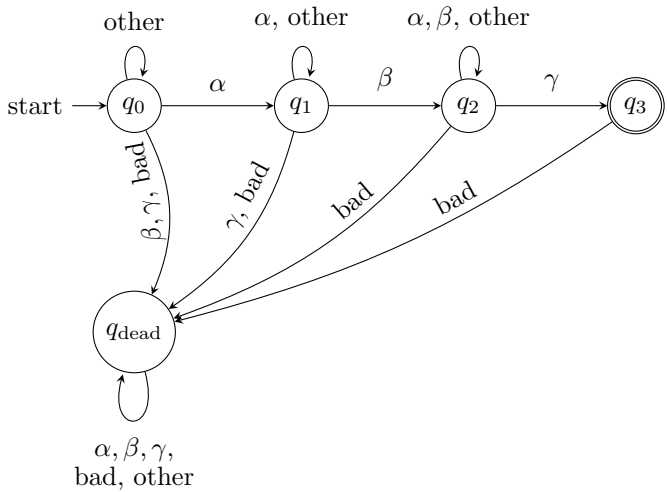

The FLTL specification is defined as
\begin{equation}\label{eq:LTL}
    \scalebox{0.85}{$F(s27\lor s34) \land G((s27\lor s34)\rightarrow F s37)\land G(s37\rightarrow F s42) \land G \neg s32.
$}
\end{equation}
This formula mandates the agent to: (i) eventually visit the region $\{s24, s34\}$; (ii) subsequently visit $s37$; (iii) finally reach $s42$; (iv) avoid the state $s32$. The corresponding deterministic finite automaton is then visualized in Fig. \ref{fig:LTL_automaton}.

The one-step cost for sum-aggregation (secondary objective) is uniformly 1. The planning horizon is set to $H = 2000$. The resulting action sequence is $(D, D, D, D, D, D,\\ D, D, D, R, R, R, R, R, R, R, U, R, U, U, D, D, D, D, D, R,\\ R, U, R, U, U)$, and as shown in Fig. \ref{Fig:fig2}, the resulting state sequence is $(s10, s11, s12, s11,s10, s3, s10, s11, s12, s13,\\ s20, s23, s20, s23, s20, s23, s29, s23, s29, s28, s27, s26, s27,\\ s28, s29, s30, s37, s44, s37, s44, s43, s42)$. The agent successfully navigates to the intermediate goals in the correct order defined by the FLTL formula. It is consistent with the first example that the agent chooses the path through $s23$ to minimize the primary max-aggregated objective. The resulting trajectory validates the effectiveness of the proposed product MDP formulation and the lexicographic cost optimization scheme in handling complex high-level tasks.

\section{Conclusion}\label{sec:conclusion}
We presented a unified framework for stochastic shortest path planning that considers mixed max–sum cost aggregations under FLTL specifications. We showed that max-aggregation breaks the standard Bellman recursion and addressed this challenge through an augmented MDP formulation. Moreover, we introduced a finite-horizon structure that avoids cyclic behaviors inherent in infinite-horizon max-aggregation. Afterwards, we combine this formulation with a product MDP and a lexicographic value iteration scheme to deal with multiple objectives that are ordered lexicographically. We show the effectiveness of our proposed framework through an example in a stochastic gridworld.

\section*{DECLARATION OF GENERATIVE AI AND AI-ASSISTED TECHNOLOGIES IN THE WRITING PROCESS}
During the preparation of this work the authors used ChatGPT in order to check grammar and refine the language. After using this tool/service, the authors reviewed and edited the content as needed and takes full responsibility for the content of the publication.

\bibliography{ifacconf}             
                                                   







\end{document}